\documentclass{jfm}
\usepackage{lineno}

\usepackage{graphicx}
\usepackage{newtxtext}
\usepackage{newtxmath}
\usepackage{natbib}
\usepackage{hyperref}
\usepackage{xcolor}

\hypersetup{
    colorlinks = true,
    urlcolor   = blue,
    citecolor  = black,
}

\newcommand{\RomanNumeralCaps}[1]
\linenumbers

\title{Vortex Dynamics During Pinch-off of Micro-Droplets}

\author{Siddhant Jain\aff{1},
  Saini Jatin Rao\aff{1}, 
  Shubhadeep Mandal\aff{1},
  Cameron Tropea\aff{2}
 \and Saptarshi Basu\aff{1}
\corresp{\email{sbasu@iisc.ac.in}}}
\affiliation{\aff{1}Department of Mechanical Engineering, Indian Institute of Science,
Bengaluru, India
\aff{2}Institute of Fluid Mechanics and Aerodynamics, Technical University of Darmstadt, 64287 Darmstadt, Germany
}

\begin{document}
\maketitle

\begin{abstract}
  Micro-droplets are extensively used in chemical, biological, and medical research, primarily for conducting various tests on samples, including living organisms, using a microfluidic framework. Recent studies have shown that the physiology of bacteria can be significantly altered when subjected to shear and/or extensional stresses. With this motivation, we perform experiments to understand the vortex dynamics involved during the pinch-off process in a cross flow droplet generator, using particle image velocimetry (PIV) to visualize the vortical structures and to quantitatively measure the associated stresses developed inside droplets. The process of pinching off inherently leads to  bi-directional acceleration of fluid  in the rapidly thinning capillary bridge, resulting in a vortex in the separated droplet as well as in the retracting ligament. We propose scaling laws for the vortical flow inside the droplet post pinch-off and predict the maximum circulation production inside droplet. Further, we discuss the vortex dynamics inside the droplet, the retracting ligament and the advancing ligament and examine the stress fields associated with this transient phenomenon.

\end{abstract}

\begin{keywords}
Droplets, Microfluidics, Vortex Dynamics, Shear Stress
\end{keywords}

\section{Introduction}
\label{sec:headings}

 Droplet microfluidics has been widely exploited in the field of applied sciences, because of the innate advantages it offers, like the  precise control over the payloads, high throughput and scalability in both space and time \citep{Moragues2023}. Biological and chemical research has specifically benefited in areas like cell sorting and single-cell analysis \textcolor{black}{through encapsulation} \citep{Mazutis2013}, synthesizing nano-materials \citep{Gimondi2023}, three-dimensional cell culture \citep{Wang}, tissue engineering \citep{Petit-Pierre2017}. Droplet generation at the micro level comes with other advantages, for example simple manipulation in channel design can lead to rapid mixing of chemical payloads inside the droplet along with providing a habitable confinement for bacterial species. This has been widely utilized in the field of on-chip drug testing \citep{Li} and for combating antimicrobial resistance \citep{Hsieh2022}. 

 The formation of droplets in flow-focusing \textcolor{black}{micro} droplet generators, as used in the present study, has been studied extensively to understand fundamental aspects of droplet generation \citep{Anna, Garstecki, Funfschilling}, flow dynamics inside droplets \citep{Ma, Roberts} and the effect of surfactants on the droplet formation process \citep{Kalli2023}. \textcolor{black}{The pinch-off of droplets is governed by the temporal interplay among viscous, inertia and capillary forces during the drop formation, a topic which remains a subject of active research interest \citep{Castrejon-Pita2015, Jiang2020, Stone_PRL, Chagot_PRF}.} It is well known that droplet microfluidics inherently involves circulation of fluid inside the formed droplets under steady conditions (when the droplet achieves a stable shape \textcolor{black}{post pinch-off}) \citep{Roberts, Ma}. This circulation enhances the mixing process when encapsulating various reagents \citep{Wang}. However, the droplet formation process involves vortical structures during pinch-off in both the generated droplet and the retracting ligament, which is known, but not yet investigated in detail.


 \textcolor{black}{ \citet{Ma} discussed the flow dynamics inside droplets during the steady state and highlighted the role of viscosity, capillary number and the droplet geometry. They concluded that the viscosity ratio is the most important factor influencing the flow topology inside the droplets. For example, they suggested a higher dispersed-to-continuous phase viscosity ratio yields lower internal circulation. \citet{Roberts} examined the velocity field inside the droplets in Hele-Shaw channels with a focus on understanding the variation in flow physics for different droplet diameter-to-channel height ratios (0.2 - 2). They associated the generation of the circulation pattern inside the droplet to the various forces (drag) acting on the droplet due to the relative flow of the continuous fluid around the droplet. The drag force led to a larger reduction in the droplet relative velocity as its size approached the channel height. They used numerical simulation to identify the change in the vortex plane when the droplet changes from fast to slower relative speeds. \textcolor{black}{They reported that the vortex forms in the $z-x$ plane (as defined in figure~\ref{fig:1}(a)) for droplet diameter to channel height $<1$ when the droplet velocity is less than half the maximum channel velocity}}. \citet{Vagner} studied the vortex dynamics inside droplets in a co-flow device and isolated the swirling and shearing contributions during droplet growth, neck formation and pinch-off phases. \textcolor{black}{They provided data on vortex dynamics and reported scaling laws for the vorticity field, albeit with limited physical insight. Furthermore, discussions about vorticity fields inside the retracting and growing ligament were not included. Recently, \citet{Shen2024} carried out similar experiments to the present study, using an altered T-shaped microchannel to highlight the evolution of micro vortices during pinch-off. They confirmed the occurrence of a high-speed jet flow leading to the generation of a counter rotating vortex pair inside the droplet. They reported that these vortices exhibited high values of vorticity compared to the high speed jet area that was dominated by strain. Although they  depicted the generation of vortices during pinch-off, scaling laws/models or further in-depth discussion of the phenomenon was not presented.}

\textcolor{black}{Micro-droplet based frameworks have evolved into a powerful tool for various bio-chemical analyses, like single cell encapsulation and screening, cytometry, synthesis of micro-particles \citep{Moragues2023}. The present study makes multiple reference to a process known as single/multiple cell encapsulation, which is widely used to analyze cells in a micro-environment like micro-droplets. For examples, if we want to study the effect of antibiotics on bacterial cells, we consider the bacterial fluid as the dispersed phase and the continuous fluid can be a bio-compatible oil (references to this methodology can be found in \citet{Li2023}). This is analogous to the situation studied in the present work. \textcolor{black}{Encapsulating single cell from a sparsely populated bacterial suspension ensures that active stresses remains unimportant, a phenomenon attributed to the collective dynamics of cells \citep{Lushi2014}.}} Recent studies have also highlighted how flow stresses can cause physiological changes \citep{Jain_soft_matter, Jain_Langmuir} and genetic regulations  \citep{Sanfilippo_review, Ramachandran2024} in bacteria. \textcolor{black}{Exposure to shear in a moving fluid caused a decline in the bacterial viability and an increment in their virulence \citep{Jain_soft_matter}, both of them indicate towards bacterial ability to reproduce or become potent post stressing conditions. It has been shown that the shear can cause significant changes in the genetic expression of the bacteria \citep{Ramachandran2024}.} \textcolor{black}{\citet{Brouzes2009} reported a 13\% loss in cell viability during the encapsulation process possibly due to the shear stress.} 
\textcolor{black}{\citet{Lam2019} reported that the high shear that live cells experience during the droplet generation process results in large viability loss, demonstrated in their study on pneumococcal genetic transformation in femtolitre droplets by encapsulating competent and non-competent strains. Furthermore, there have been reports \citep{Tan2022} investigating the effect of an increase in diffusion time from a smaller droplet to a larger droplet on the growth dynamics of bacterial cells. The diffusion time can be enhanced inside droplets by inducing chaotic advection. }
 
 \textcolor{black}{One of the primary areas benefiting from the present study involves cell encapsulation, manipulation and screening \citep{Ma, Ma2015_a,  Vagner} using a micro-droplet framework that essentially requires understanding the vorticity dynamics well. The topic that needs attention and forms the main focus of the present work, is how the vorticity field inside the droplet and in the remnant ligament evolves during post pinch-off. In particular, how does the stress field evolve spatially and temporally? 
 Most of the studies concerning the pinch-off dynamics exclude the inner flow dynamics that forms the main theme of the present work. Finally, an attempt is made towards generalization by developing scaling laws that to date are missing, likely due to the involvement of large number of geometric and flow parameters \citep{Baroud}.} 
 

We discuss experimental findings on vortex dynamics for pre, post and during the pinch-off process for a flow focusing device that is one of the most widely used droplet generator \citep{Moragues2023},  to shed light on the shear and elongational stresses developed due to different vortical structures. Droplet generation for a range of capillary number ($Ca_\mathrm c$ = $\mu_{\mathrm c}u_\mathrm {c}/\sigma$, where subscripts \textit{c} denote properties for continuous fluid and subscript \textit{d} denotes for dispersed fluid, $\mu$ the dynamic viscosity, \textit{u} the velocity and $\sigma$ the surface tension) from 0.008 to 0.02 for a viscosity ratio ($\nu_{d}/\nu _c$, $\nu$ being the kinematic viscosity) of 0.02 is investigated. \textcolor{black}{We present most of our data in terms of the parameter $\varphi$, defined as the ratio of flow rates, $Q_\mathrm {c}$ to $Q_\mathrm {d}$, which is varied from 4 to 10 (except for droplet diameter data, that has been presented for a range of 4 till 12) by varying $Q_\mathrm {c}$.} We investigate one cycle of droplet generation starting from a pinch-off until the next pinch-off. When the pinch-off occurs, it results in  a post pinch-off vortex in the droplet, and a vortex in the retracting ligament, as shown in figure~\ref{fig:1}. Once the retraction of the interface stops, the advancing ligament also exhibits a vortex.
We do not discuss the steady-state condition of the droplet that occurs post pinch-off and once the droplet reaches its final shape and relative velocity.

The paper is organized as follows: Section II contains details about the experimental setup and measurements, section III discusses the results about the droplet formation physics and its characterization, followed by pre and post pinch-off vortex and stress dynamics. Finally, conclusions of the study are presented in section IV.

\section {Experimental Setup and Measurements}\label{exp setup}
\begin{figure}
  \centerline{\includegraphics[width=1\columnwidth]{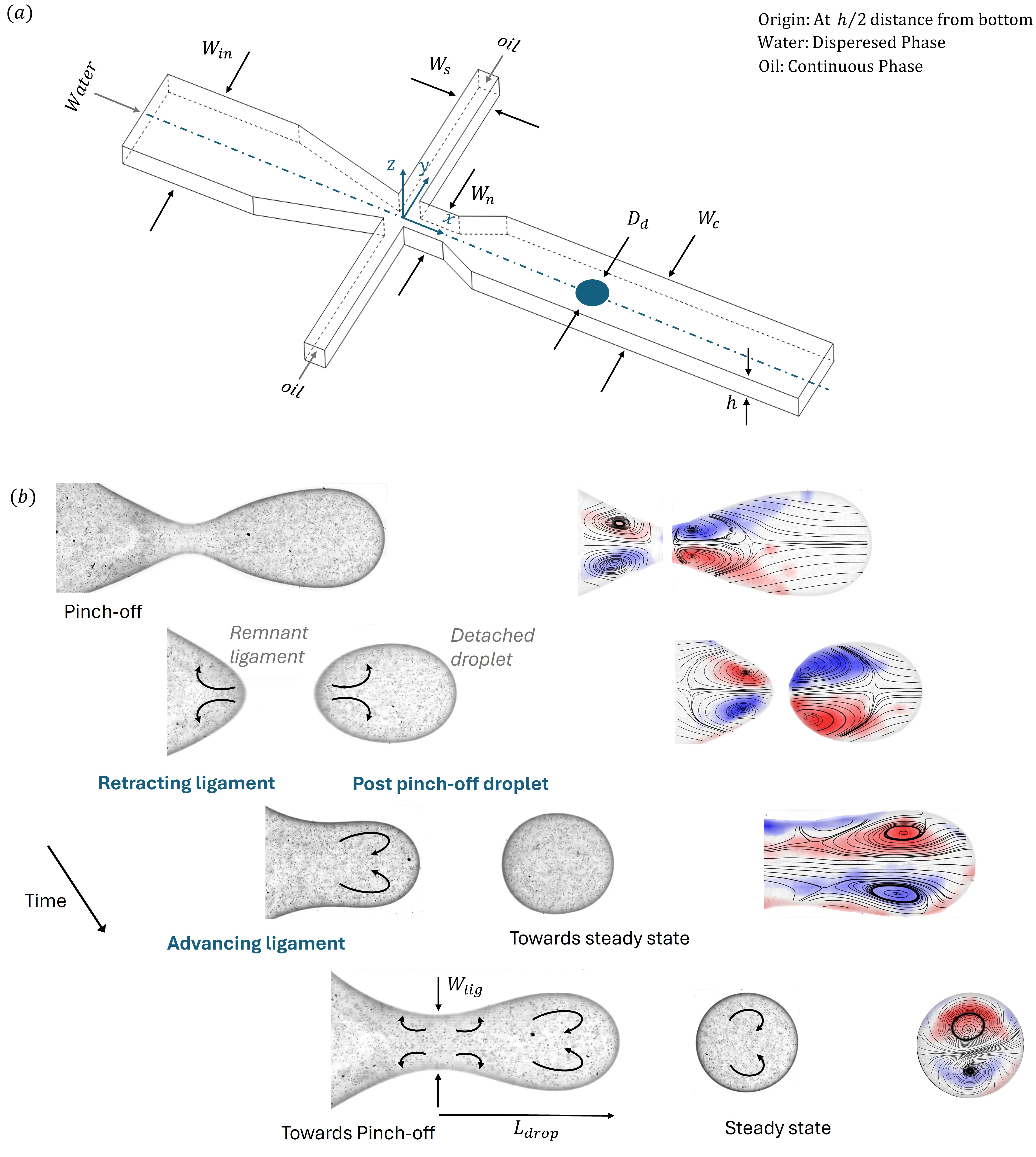}}
  \caption{(a) Three-dimensional schematic of the droplet generator device with associated nomenclature (not to scale) (b) Raw images and corresponding streamlines shown for post pinch-off vortex, retracting ligament vortex and advancing ligament vortex. For completeness, the streamlines in the steady state droplet are also shown in the last row. The streamlines are generated (for visualization purpose) by subtracting the average velocity inside the droplet as discussed in \citet{Vagner}, and then they are superimposed on the vorticity contours.}
\label{fig:1}
\end{figure}

The schematic for the microfluidic chip is shown in figure~\ref{fig:1}(a). The microfluidic channel is fabricated using polydimethyl siloxane (PDMS) following  standard lithography, as discussed in \citet{Jain_soft_matter}. \textcolor{black}{The channel width-to-height $(W_c/h = 270/90$ with $\pm$ 8 $\mu m$ deviation in each of the dimensions resulting from the fabrication process) ratio of the droplet generator is kept around three to ensure that the dominant flow dynamics occur in the observation plane with the neck width ($W_n$) being 200 $\mu m$.} The nomenclature is shown in figure~\ref{fig:1}(a). The drop geometry is defined using the ratio $D_d/h$ ($D_\mathrm {d}$, i.e., the final droplet size at steady state). Two syringe pumps are used for pumping the continuous (silicone oil, $\nu_\mathrm {c}$ = 50 cst) and dispersed phase (water, $\nu_\mathrm {d}$ = 1 cst) at different flow rates, regulating $\varphi$  between 4 and 12. \textcolor{black}{The ratio of droplet diameter to \textit{h} of the channel, varied from 2.3 - 1.5, with increasing $\varphi$ resulting in an ellipsoidal drop geometry.} The value of $\sigma$ for the oil/water interface is 25 mN/m, measured using the standard pendant drop method. The value of Bond number (\textit{Bo} = $\Delta\rho g l^{2}/\sigma$, $\Delta\rho$ being the density difference, \textit{g} the gravitational acceleration and \textit{l} being a characteristic length ($\sim$\textit{O}($10^{-6}$))) is much less than unity ($\sim$\textit{O}($10^{-4}$)) rendering gravitational forces negligible. \textcolor{black}{A global Reynolds number ($Re$) defined as \textit{Re} = $\rho_c u_c W_c/\mu_c$ gives a value of the order of 0.01. However, being a multiphase system, we can define $Re$ based on local flow parameters during necking as will be discussed in section~\ref{sec:VD post}}

To visualize the flow inside the dispersed fluid, polystyrene particles of diameter 0.9~$\mu$m are mixed in the dispersed phase fluid (water). The experiments are conducted using a 20$\times$ objective \textcolor{black}{(Olympus LUCPlanFL N)} under an inverted microscope (Olympus CKX53) with bright-field lighting using a high-speed camera (Photron Mini UX 100) at 8000 - 25000 Hz. \textcolor{black}{This resulted in different fields of views (FOV) for different cases. For example, for lower $\varphi=$ 4 and 6, the recording rate was 10240~Hz (FOV $\sim$ 0.22 mm x 0.58 mm), whereas for higher $\varphi$ the rate was  12500~Hz (FOV $\sim$ 0.18~mm x 0.58~mm). For capturing the ligament thinning phenomenon, the recording rate was kept at 25000~Hz that resulted in a FOV of 0.09~mm x 0.58~mm. A higher recording rate resulted in lower FOV, but this was employed for higher $\varphi$ that generated droplets with smaller diameter, thereby supporting the experiments.}   The exposure time is kept short to avoid motion blur. A key challenge when experimenting with a transient problem spanning such a wide range of time and velocity scales is determining the appropriate frame rate for imaging. During pinch-off, the particles near the neck accelerate; however, at the far end of the droplet the particles are almost motionless. However, it is also seen that the dynamics of the far end do not significantly alter the physics near the pinch-off region. \textcolor{black}{The peak velocity ($U_\mathrm{peak}$) as will be used in the sections ahead, is identified using a standard particle tracking methodology.} All the experiments are conducted in the dripping regime of the generator. The range of $Ca_\mathrm {c}$ considered here generated droplets with only final circular shapes (in the $y$ plane), i.e., plug-shaped droplets are not studied. 

\textcolor{black}{The captured images are calibrated using a glass calibration plate with consecutive markings at 200~$\mu m$ spacing.} The images are initially processed using the open source toolbox ImageJ which are then processed using PIVlab software \citep{PIVlab} to obtain the velocity field inside the droplet. The consecutive images are cross-correlated using multi-pass processing with decreasing interrogation window size from 64x64~px to 32x32~px \textcolor{black}{with an overlap of 50\%}. A pixel resolution of $\sim$ 2200~px/mm is achieved \textcolor{black}{with} a distance of $\sim 7~\mu m$ between consecutive vectors \citep{Leong2016} \textcolor{black}{(as shown in figure S1 in the supplementary data). In our study, we observed an average pixel displacement of 6 pixels between two frames. This leads to a global random error of 0.28 pixel or 4.7\% \citep{Raffel2018}. The error estimation using the synthetic images in the PIV Lab gives a very high correlation coefficient (0.99) with a bias error within 0.1\% \citep{Chandel2021}. For our experiments, we achieve a global correlation coefficient of 0.95.}

The vorticity (\(\omega=\frac{\partial v}{\partial x}-\frac{\partial u}{\partial y}\), \textit{u} and \textit{v} being velocity in \textit{x} and \textit{y} direction respectively)  represents the core vorticity of the vortex \textcolor{black}{identified using the $\lambda_\mathrm {ci}$ method \citep{ZHOU1999}, that computes the eigenvalue conjugate pair of the velocity gradient tensor. }
\textcolor{black}{We define the velocity gradient tensor as}

\textcolor{black}{\begin{equation}
    \nabla u = 
\begin{bmatrix} \partial u/\partial x & \partial u/\partial y \\ \partial v/\partial x & \partial v/\partial y 
\end{bmatrix}
\end{equation}}
\textcolor{black}{Then, $\lambda$ represents the complex conjugate eigenvalues of $\nabla u$ that can be decomposed as $\lambda_r + i\lambda_{ci}$. The condition $\lambda_{ci} > 0$ then indicates that local swirling exists. The vortex core is finally identified by the connected region where the values of $\lambda_\mathrm {ci}$ are positive \citep{ZHOU1999,Rao_PRF} (as shown in figure S2 in the supplementary data).}
For all experimental data presented (except for contour plots in figures~\ref{fig:3} and ~\ref{fig:6}), the absolute values of $\omega$ are considered. \textcolor{black}{The values of $\omega$ are averaged in space for the core region identified using the $\lambda_\mathrm {ci}$ method.}  For a single droplet,  $\omega^+$ and $\omega^-$ above and below the $x$ axis are averaged and the same is done for the corresponding core circulation ($\Gamma)$, calculated as \textcolor{black}{\(\int_c|\omega| dA \), where \textit{c} represents the core region of the vortex. The implementation of the area integral involves summing over the product of vorticity value and area of each pixel within the core area}.  The in-plane shear rate ($\epsilon$) is defined as  \((\epsilon = \frac{\partial v}{\partial x}+\frac{\partial u}{\partial y})\) and the extension rates in the \textit{x} and \textit{y} directions are defined as \(\eta_\mathrm x=\partial u/\partial x\) \ and  \(\eta_\mathrm y=\partial v/\partial y\) respectively. Due to the transient nature of the droplet formation, it is difficult to identify an appropriate time scale; hence, vorticity and stress values are multiplied by the capillary time scale defined as:
\begin{equation}
t_{\mathrm {cap}}=\sqrt{\frac{\rho_\mathrm cW_\mathrm n^{3}}{\sigma}}    
\end{equation}
 to render them dimensionless. \textcolor{black}{In figure~\ref{fig:2}(c) the variation of pinch-off time ($t_\mathrm{pinch}$) with the $Ca$ is depicted where the $t_\mathrm{pinch}$ is calculated by first fixing the pinch-off point followed by the time taken for the pinch-off from an instance when $W_\mathrm{lig}^*$ $\approx$ 0.4 ($W_\mathrm{lig}^*$ is the width of the ligament which is rendered non-dimensional using $W_n$ ) along the pinch-off point.} For figure~\ref{fig:2}(d and e) the dimensionless time is represented as $t^*=t/t_\mathrm{cap}$ where, $t^*=0$ corresponds to $W_\mathrm{lig}^* \approx 0.4$ . For all other plots presented, the dimensionless time is defined as $T^*=(t-t_\mathrm{pinch})/t_\mathrm{cap}$, where $T^*=0$ denotes the pinch-off point. The inertia-capillary time scale, as defined above, yields a time scale of order $10^{-4}$~s. The same order has been observed for the phenomenon in the experiments as well. It has been reported that in the vicinity of the pinch-off, all three forces, i.e., inertial, viscous and capillary forces balance and no two forces dominate, unlike in the earlier stages of the phenomenon \citep{Castrejon-Pita2015, Jiang2020}. Moreover, the thinning of the ligament results in accelerating fluid out of the neck (as will be discussed in section~\ref{sec:VD post}), a phenomenon attributed to the fluid inertia \citep{Castrejon-Pita2015}. Hence, the inertia-capillary time scale is the most appropriate scale for rendering the time dimensionless. Further relevance of this choice will be addressed later. Other relevant scales are mentioned during the upcoming discussions, wherever necessary.

\textcolor{black}{We used three channels derived from a same master mold, resulting in only minor differences in the dimensions of the PDMS channels. For the pinch-off data, two runs were taken from each channel and were then averaged to obtain the data presented (a total of 6 runs). To obtain the diameter (in steady state), droplets at two different locations were considered inside the channel, resulting in six data points.  The droplet generation process was found to be highly repeatable with minimal deviation observed when employing fresh channels derived from the same lithography-based master mold.}

\section {Results and Discussion}
\subsection {Droplet Formation and Characterization}{\label{section3.1}}

\begin{figure}
  \centerline{\includegraphics[width=1\columnwidth]{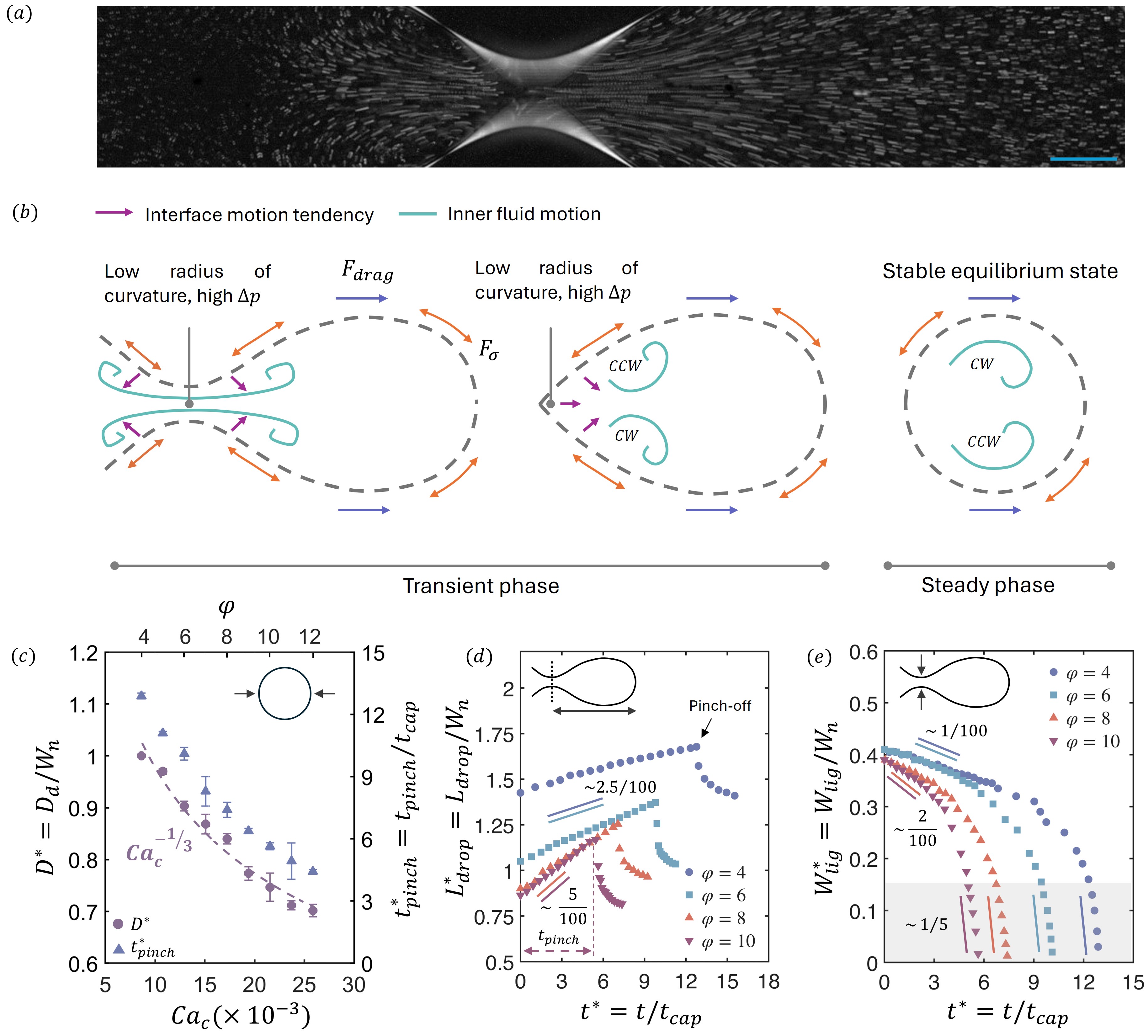}}
  \caption{(a) \textcolor{black}{A long exposure image of the ligament thinning process. The scale bar represents 40 $\mu m$} (b) Schematic depicting forces and interfacial motion involved during the droplet formation process. $F_\mathrm {drag}$ represents a combination of shear and pressure gradients in the continuous fluid and $F_\mathrm\sigma$ is the interfacial tension force. $\Delta p$ refers to the Laplace pressure difference across the interface \textcolor{black}{(c) The left and right y-axis depicts the variation of $D^*$ and $t_\mathrm{pinch}$ respectively with $Ca_\mathrm {c}$ and $\varphi$} (d) Time evolution of dimensionless droplet length $L_\mathrm {drop}^*$. Reference time ($t^*=0$) corresponds to $W_\mathrm{lig}^* \approx 0.4$ (e) Time evolution of the ligament width ($W_\mathrm {lig}$) during necking. The gray zone in (e) indicates the time when the slope of $W_\mathrm {lig}^*$ becomes constant and similar for each of the $\varphi$ values.}
\label{fig:2}
\end{figure}

\textcolor{black}{We first discuss trends related to droplet generation including how the pinch-off process influences the final droplet size}. Figure~\ref{fig:2}(a) depicts the thinning of the ligament through a long exposure image (created by projecting several frames pre pinch-off) where the region near the neck has longer streaks compared to other regions, indicating larger velocity. Figure~\ref{fig:2}(b) illustrates the forces that act on a droplet in various stages and the directional tendency of interface movement. In the dripping regime (in which the experiments are conducted), the dispersed phase is initially convected with the continuous phase fluid, resulting in its stretching. A blunt nose is formed at the tip of the dispersed fluid (see figure~\ref{fig:1}(b)), which is subjected to shear drag and pressure build-up from the channel restrictions \citep{Kalli2023}. As the droplet expands, the drag force on the droplet increases \citep{Roumpea2019}, and a reversal of flow is observed at the necking zone (figure~\ref{fig:2}(a),(b)), marking the beginning of necking. Then the interfacial tension begins to pull the interface backward (figure~\ref{fig:2}(b)). Finally, the cylindrical thread is subjected to the capillary (Rayleigh-Plateau) instability, resulting in the pinch-off \citep{Baroud}.

Figure~\ref{fig:2}(c) shows the variation in the dimensionless droplet diameter ($D^{*}=D_\mathrm d/W_\mathrm n$) and $t_\mathrm{pinch}$ with increasing $Ca_\mathrm c$ and $\varphi$. As has been found in previous studies \citep{Nie2008}, increasing  $Ca_\mathrm c$ results in a decrease in diameter. For the present device and conditions, $D^{*}$ is seen to decrease as $Ca_c^{-1/3}$ \textcolor{black}{(similar to what was observed by \citet{lee_role_2009})}. \textcolor{black}{This decrease in droplet diameter is due to the fact that with an increase in $Q_\mathrm{c}$ or $\varphi$, the shearing effects increase, leading to more rapid thinning of the neck. A faster thinning implies less volume accumulation and hence smaller droplets. The decrease in $t_\mathrm{pinch}$ (figure~\ref{fig:2}(c, d)) with increasing $\varphi$ is consistent with this explanation.} \textcolor{black}{It should to be noted that the dimensionless droplet length $L^*_\mathrm{drop}(=L_\mathrm{drop}/W_\mathrm{n})$ in ~\ref{fig:2}(d) is plotted against $t^*$ that is defined from the point where the dimensionless ligament width $W_\mathrm{lig}^*$ $\approx 0.4$ (figure~\ref{fig:2}(e)). This reference ($W_\mathrm{lig}^*\approx 0.4$) comes from an experimental limitation i.e., smaller field of view when recording at 25000 Hz to capture the ligament dynamics as mentioned in section~\ref{exp setup}.}

\textcolor{black}{Before pinch-off, $L^*_\mathrm{drop}$ for $\varphi=$ 4 and 6 exhibits a \textcolor{black}{similar} slope whereas the higher $\varphi$ cases (8 and 10) displays a sharper slope that increases monotonically with $\varphi$.} Post pinch-off, the droplet length decreases rapidly toward a fixed length in the stable state. \textcolor{black}{This is primarily because of the surface tension of the deformed pinch-off droplet.} \textcolor{black}{Further, it can be seen from figure~\ref{fig:2}(e) that the $W_\mathrm{lig}^*$ initially decreases linearly with different slopes for smaller $\varphi$ (4 and 6) and larger $\varphi$ (8 and 10). A change in the slope is then observed, followed by the final decreasing slope that is identical for all $\varphi$}, suggesting a capillary dominated break-up, independent of the value of $Q_\mathrm c$. Hence, the pinch-off can be understood as comprising two stages, the initial stage is dependent on $\varphi$ i.e., the drag, and the second stage is capillary dominated. \textcolor{black}{Interestingly, even in turbulent conditions it is seen that the final breakup process is capillary dominated (\citet{EASTWOOD2004, Zhong2024}). In the literature, ligament thinning is presented in terms of the minimum neck width that exhibits power-law relation in time for various regimes \citep{Castrejon-Pita2015}. However, we do not provide such arguments to align with the current focus, i.e., investigating the generation of vorticity due to pinch-off rather than pinch-off itself.}

 \subsection {Vortex Dynamics: Post Pinch-Off Vortex}\label{sec:VD post}

\begin{figure}
  \centerline{\includegraphics[width=1\columnwidth]{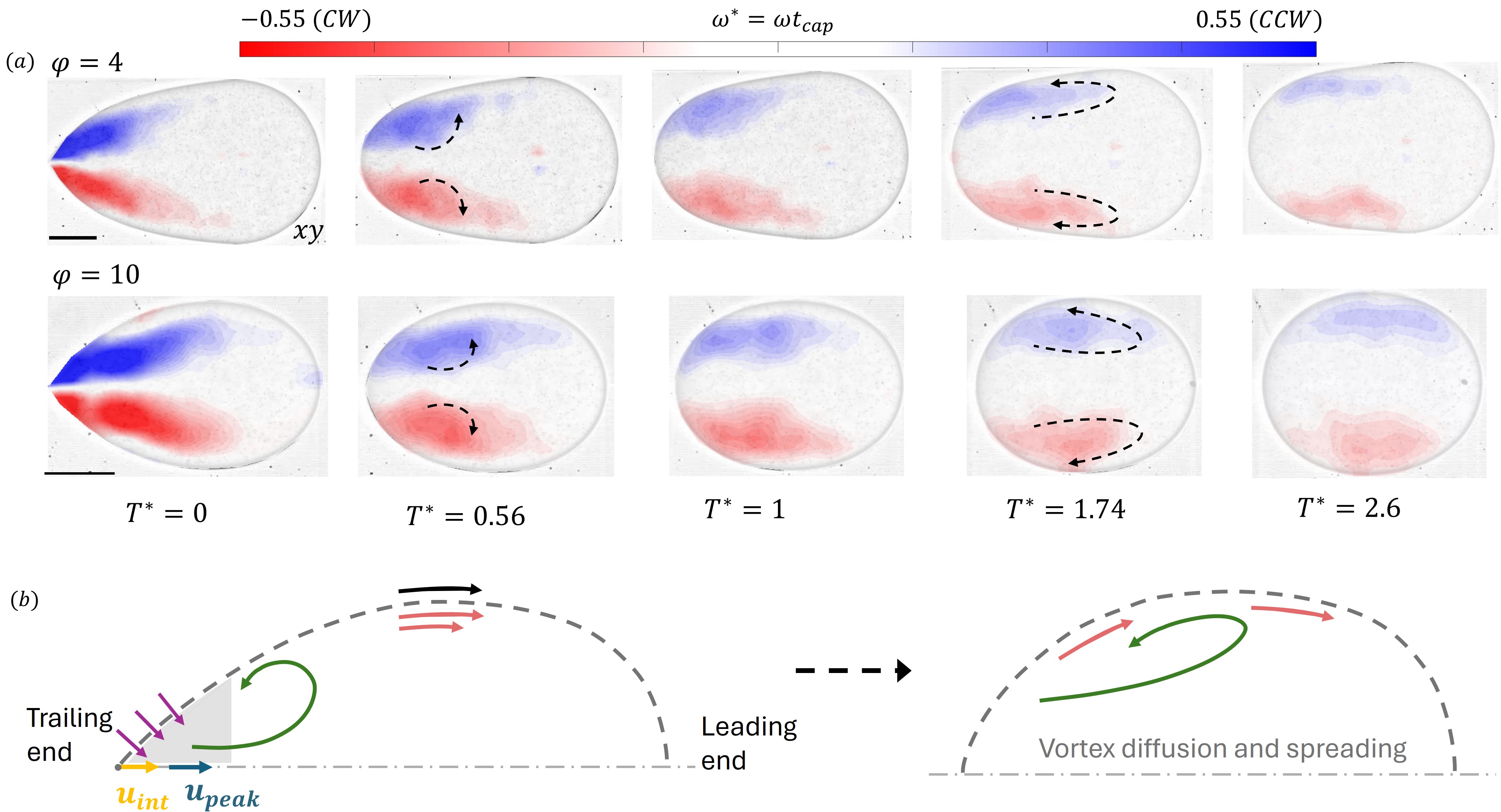}}
  \caption{(a) Time evolution of $\omega^*$ for $\varphi = 4$ and $\varphi = 10$. The scale bar represents 50~$\mu m$ (b) Decay mechanism of the post pinch-off vortex during transient state when the droplet shape is unstable. The purple colored arrows on the interface depict the tendency of interface motion and the gray zone represents the region being deformed. \textcolor{black}{The yellow vector represents the velocity of the trailing end of the drop interface $(U_\mathrm{int})$ and the blue vector represent the peak velocity inside the droplet which is achieved generally near the trailing end.}}
\label{fig:3}
\end{figure}

The necking phenomenon results in a bi-directional acceleration of the fluid being evacuated out of the thinning ligament \citep{Castrejon-Pita2015} as shown in Figure~\ref{fig:2}(a). The fluid accelerating in the forward direction flows into the droplet, and the maximum velocity of the fluid is reached at the time of pinch-off. This fluid accelerates into a comparatively quiescent droplet resulting in a vortex formation, as is schematically depicted in figure~\ref{fig:2}(b) and experimentally observed in figure~\ref{fig:3}(a). This strong fluid acceleration is due to rapid neck contraction leading to pinch-off and the very low radius of curvature at the remnant neck thereafter, which results in a high Laplace pressure (figure~\ref{fig:2}(b)) imparting a strong internal flow into the droplet. The resulting vortex remains confined at the trailing tip of the droplet during pinch-off and dies out as the droplet approaches an equilibrium shape. We now focus on this post pinch-off vortex and the transient vortex dynamics of its decay.

The pinch-off of the droplet simultaneously disrupts the supply of vorticity. Subsequent to pinch-off, the pointed tail (i.e., the trailing end) of the droplet rapidly retracts, leading to a sharp decrease in droplet length (see figure~\ref{fig:2}(d)) and a more circular shape, approaching that of the final steady state form.  
The vortex created in the tail of the droplet during the pinch-off process starts to decay as soon as the pinch-off occurs, as can be seen in figure~\ref{fig:3}(a). This decay occurs over a time scale $ T^*\sim 2.3$, which is much shorter than the time taken for the droplet to minimize its surface energy and become stable in a circular shape. The length of droplet is much larger for $\varphi = 4$ (can be deduced from figure~\ref{fig:2}(d)) in which the vorticity zone is spread over a smaller area compared to $\varphi = 10$, where the area covered is more than 50\% of the droplet size. 
It is worth noting that most studies dealing with mixing inside micro droplets ignore this vortex and study only the steady state vortex (as shown in figure~\ref{fig:1}(b)), which exhibits vorticity in the opposite sense after the decay of the post pinch-off vortex.

\begin{figure}
  \centerline{\includegraphics[width=01\columnwidth]{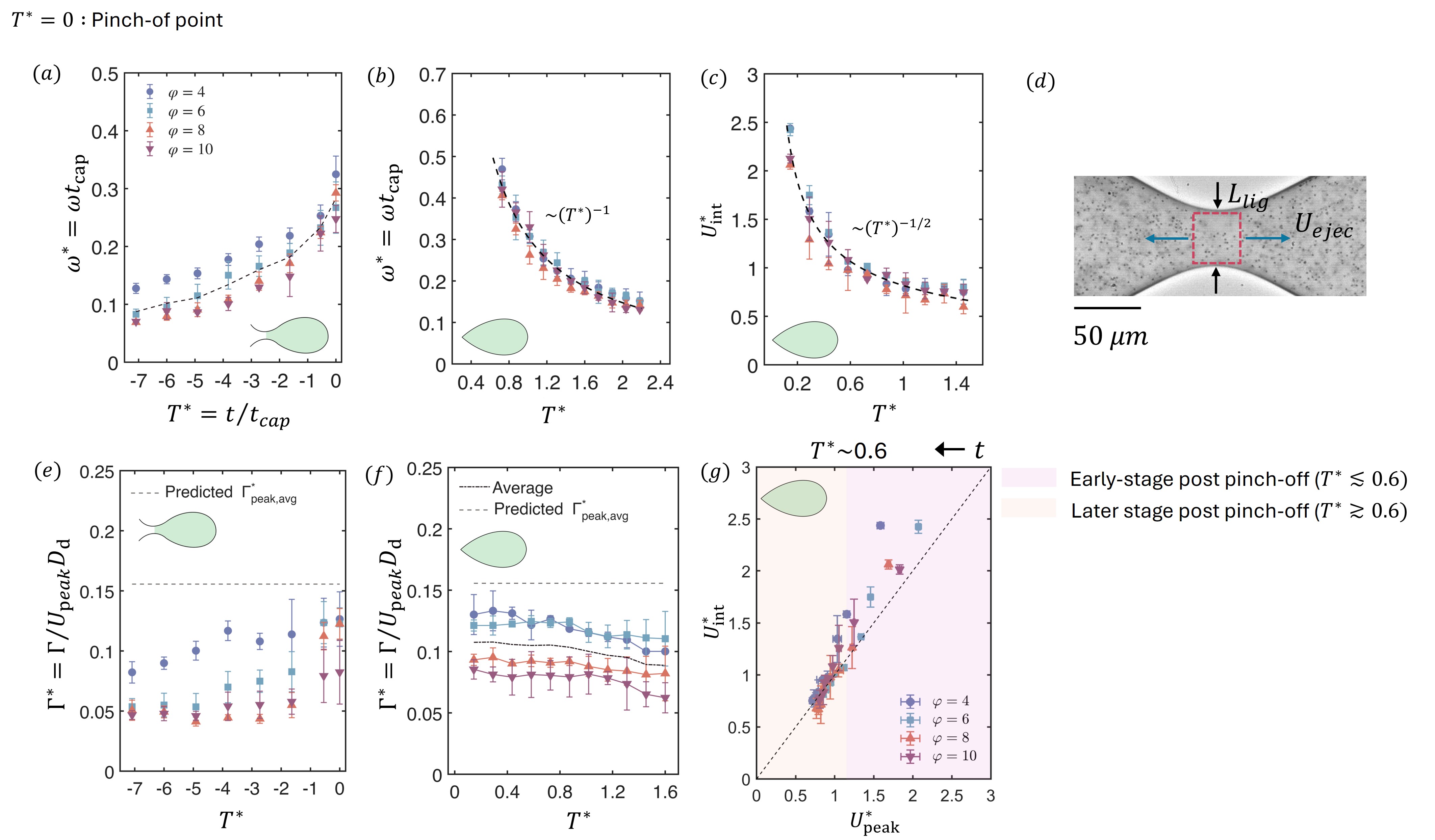}}
  \caption{(a) Evolution of $\omega^*$ during the necking phase. The green marked zone inside the droplet depicts the area over which $\omega^*$ is evaluated. (b) Evolution of $\omega^*$ post pinch-off. The dotted line represents a fitted curve (c) Evolution of $U^*_\mathrm {int}=U_\mathrm {int}(t_\mathrm {pinch}/D_\mathrm d)$ post pinch-off (d) Control volume for the model presented in equation~(\ref{Equation:3.4}) (e) Evolution of $\Gamma^*$ during necking phase. The dashed line represents the peak value of $\Gamma^*$ presented as the average of each case using equation~(\ref{Equation:3.4}). $U_\mathrm {p,max}$ represents the maximum value of the peak velocities for each $\varphi$ post pinch-off. (f) Evolution of $\Gamma^*$ post pinch-off. The dot-dashed line represents the average of each case. \textcolor{black}{(g) The trailing end interface velocity ($U_\mathrm {int}^*$) plotted against the peak velocity inside the pinched-off droplet \textcolor{black}{($U_\mathrm {peak}^*=U_\mathrm{peak}(t_\mathrm {pinch}/D_\mathrm d)$).}}}
  
\label{fig:4}
\end{figure}

Figure~\ref{fig:4}(a) quantitatively shows the evolution of vorticity in the droplet before and after pinch-off. For all values of $\varphi$, $\omega^*(=\omega t_\mathrm {cap})$ is seen to increase until the pinch-off time, after which it decays with time as $\omega^* \sim (T^*)^{-1}$, with an accommodated time lag of $T_\mathrm o^* (\sim O(10^{-4}s))$.  The vorticity is decreased by the continuous phase fluid acting in the opposite direction at the interface. This causes the vorticity to decay in magnitude as depicted in figure~\ref{fig:4}(b). \textcolor{black}{We observe minimal variation in the vorticity values on changing $\varphi$ (figure~\ref{fig:4}(b)). This is because the vorticity generation is primarily a result of the final thinning of the ligament that exhibits a similar rate of contraction (figure~\ref{fig:2}(e)), thereby making the vorticity values independent of the variation in $\varphi$. This is governed by the capillary time scale, which is a constant for the chosen set of parameters. Further discussion on this mechanism is made later in this section.}

The vortex formed in the post pinch-off droplet is symmetric about the \textit{x-}axis (figure~\ref{fig:3}). Further, owing to the geometry of the channel, we assume the three-dimensional effects to be minimum; hence, stretching of the vortex structure along its vorticity vector is expected to be insignificant. If we scale the core vorticity with the velocity gradients as
\begin{equation}
    \omega ~\sim \frac{U_\mathrm r}{L_\mathrm r}
    \label{Equation:3.1}
\end{equation}

 \noindent where \textit{r} as subscript refers to the \textit{reference scale},  then $U_r$ should scale with the velocity of the trailing end velocity ($U_\mathrm {int}$) and the length scale $L_\mathrm r$ will be the viscous length scale i.e., $L_\mathrm r \sim t^{1/2}$. This is because at the interface, a slip condition is expected, leading to a finite tangential interface velocity. \textcolor{black}{The imposition of the shear from the continuous fluid creates a region with localized velocity gradients near the interface and a jump in velocity gradient is expected across the interface which depends on the viscosity ratio \citep{Ma}}. From the shear matching condition at the interface we have,
 
 \begin{equation}
 \mu_\mathrm c\frac{\partial u}{\partial y}|_\mathrm {out}=\mu_\mathrm d\frac{\partial u}{\partial y}|_\mathrm {in}
 \label{Equation:3.2},      
 \end{equation}

 \noindent where subscript \textit{`out'} refers to continuous fluid and \textit{`in'} refers to the dispersed fluid. From equation~(\ref{Equation:3.2}), it can be deduced that the velocity gradient inside the dispersed phase droplet will be much larger than the continuous phase gradient. The magnitude of the inner velocity gradient decreases with distance from the interface (depicted in figure~\ref{fig:3}(b)) and is further decreased by the vortex rotating in the opposite sense. Moreover, the pronounced deformation of the interface at the trailing end post pinch-off will create vorticity because of interface motion. This vorticity must then diffuse into the interior of the droplet \citep{Song1999-kh}. While the specific role of vorticity generated by the interfacial motion could not be definitively identified, observations during the decay phase show the vortex moving away from the trailing end, diffusing inside the droplet, as schematically shown in figure~\ref{fig:3}(b). Although capturing velocity gradients near the interface is experimentally challenging, their presence is indisputable and they have a considerable impact on the vortex dynamics. The interaction of these gradients near the inner interface with the generated vortex results in its decay. It is further seen (figure~\ref{fig:4}(c)) that the decay of the trailing end interfacial velocity ($U_\mathrm {int}$) follows the scaling $U^*_\mathrm {int}\sim (T^*)^{-1/2} $. \textcolor{black}{It has been previously shown \citep{Shikhmurzaev2000} for capillary pinch-off of liquid neck that the retraction velocity of the cusp post pinch-off follows a similar scaling as observed here. The shape of the interface during pinch-off process being self-similar, imposes the tip position to scale as $t^{1/2}$ \citep{Eggers1997-de} resulting in the retraction velocity to decrease as a power of $1/2$. Physically, just after pinch-off, the trailing end of the detached droplet forms a highly curved cusp with large capillary pressure. This drives local interface deformation, smoothing the cusp and causing capillary-dominated retraction. As the interface flattens over time, curvature and driving force decrease. Consequently, the tip velocity $(U_\mathrm{int})$ gradually declines, following the stated power law}. In our case, $U^*_\mathrm {int}$ appears to be the most relevant velocity scale post pinch-off, since the capillary forces are high at the trailing end at this period, which is expected to affect $\omega^*$ inside the droplet. \textcolor{black}{Using the relation~\ref{Equation:3.1} yields the observed scaling for the core vorticity in dimensional form, i.e.,}

 \begin{equation}
    \omega ~\sim \frac{U_\mathrm r}{L_\mathrm r}\sim\frac{U_\mathrm {int}}{{(\nu t)^{1/2}}}\sim\frac{t^{-1/2}}{t^{1/2}}\sim t^{-1}
    \label{Equation:3.3}
\end{equation}

\textcolor{black}{Moreover, $U^*_\mathrm{int}$ has been rendered non-dimensional with the velocity scale $(D_\mathrm d/t_\mathrm {pinch})$ that is dependent on $\varphi$ (figure S3, in supplementary data) resulting in a collapse of all the measured data (figure~\ref{fig:4}(c)).} The scalings obtained experimentally for $\omega^*$ indicate that there is a time lag in the scaling of $\omega^*$ relative to the $U^*_{int}$ scaling that is used to explain the former. To justify this, we define a viscous time scale \(\tau_\mathrm {vis}={\mu_\mathrm d}/{\rho_\mathrm d U_\mathrm {int}^2}\) \citep{Ni_2024} for the time period of retraction just after the pinch-off. In this phase, we expect a high viscosity, high velocity condition, since the length scales are extremely small with very high capillary and viscous forces. The order of the time lag seen for the $\omega^*$ is $O(10^{-4} s)$, which is of the same order of $\tau_\mathrm {vis}$ with $U_\mathrm {int}\sim(10^{-1}  m/s)$. Hence, it can be inferred that the interface velocity is imposed on the vorticity decay with a time lag that responds to the relaxation of the interface just post pinch-off. \textcolor{black}{Moreover, the time taken for the vortex core to diffuse in the experiments was observed to be $O(10^{-4})$ sec which is of the same order as the diffusive time scale defined as $L^2/\nu$ considering the radius of the droplet as the characteristic length scale. This justifies the choice of diffusive length scale in equation~\ref{Equation:3.3}. Furthermore, it is also seen that $t_{\rm cap}$ is of the same order as the diffusive time scale i.e., $O(10^{-4})$. Hence, a `cause and effect' dynamics can be established between them.} This retraction of the trailing end acts like an impulsive force, the effect of which is seen on the \textcolor{black}{core} vorticity after the characteristic time described above. 

Furthermore, the evolution of core circulation ($\Gamma^*$) is depicted in figure~\ref{fig:4}(e, f). As expected, $\Gamma^*$ increases up to the pinch-off time ($T^*=0$). Post pinch-off (figure~\ref{fig:4}(f)), the circulation is seen to decrease weakly. $\Gamma^*$ being the area integral of vorticity, indicates that the encompassed area of the vortex is increasing, since the vorticity level is decreasing (figure~\ref{fig:4}(b)). We predict the scale of the maximum circulation ($\Gamma^*_\mathrm {peak}$) using equation~(\ref{Equation:3.4}) and represent the average of each case in figure~\ref{fig:4}(f). We consider a control volume (figure~\ref{fig:4}(d)) around the neck of the droplet and the constant decrease of its radius during the necking phase, as observed in figure~\ref{fig:2}(e), to determine the volume flux of vorticity in the axial direction. We assume that the length of the control volume follows the imposed characteristic length $W_\mathrm {lig}$. \textcolor{black}{Then the following formulation with the scaling constant as unity for the vorticity flux through a slug approximation \citep{DABIRI_GHARIB_2005} yields ($\Gamma^*_\mathrm {peak}$)}:

\begin{equation}
    \frac{d\Gamma_\mathrm s}{dt} \sim \frac{1}{2}U_\mathrm {ejec}^2(t)\sim\frac{1}{2}\left( \frac{\Omega(t)}{A(t)}\right)^2\sim\frac{1}{2}\left(\frac{\Omega^2(t)}{(\pi W_\mathrm {lig}^2/4)^2(t)}\right)\
    \label{Equation:3.4}
\end{equation}

 \vspace{1cm}
 \noindent $\Gamma_\mathrm s$ represents the circulation values predicted using the slug flow approximation, $U_\mathrm {ejec}$ is the velocity with which the accelerated fluid comes out of the necking zone in the droplet, $\Omega$ is the volumetric flux obtained using ${W_\mathrm {lig}}$ and \textit{A} is the cross-sectional area of the thinning ligament. The above formulation is applied only during the necking phase, where the rate of thinning is found to be constant, and feeding to the droplet is cut off from the upstream part. The data obtained from figure~\ref{fig:2}(e) is used to find the $\Omega$ and \textit{A(t)}. A time integral then yields the variation of $\Gamma^*$ in time and the order of $\Gamma^*_\mathrm {peak}$. It can be seen in figure~\ref{fig:4}(e,f) that the average of the maximum predicted value lies close to the maximum obtained by averaging the experimental data for each $\varphi$. \textcolor{black}{It is worth noting that the slug model that is generally an inviscid approach, predicts the maximum value to the correct order and viscosity shall appear as an additional correction term, which can be accounted for in more exact representation of the circulation production \citep{Dabiri2004, Krieg2013}. On defining a local Reynolds number at the instance we apply the slug flow, we get}
 

\textcolor{black}{\begin{equation}
Re_{\rm slug} = \frac{\rho_\mathrm dU_\mathrm{ejec}W_\mathrm{lig}}{\mu_\mathrm d} \approx 3.4
\label{Equation:3.5}
\end{equation}}
\textcolor{black}{The length scale is taken at the point where we start considering the values of $W_{lig}$ and $U_\mathrm{ejec}$ is the velocity of particle near the thinning zone in the ligament found using particle tracking methodology (as discussed in section~\ref{exp setup}), which was found to be of same order until the pinch-off. \textcolor{black}{As can be seen, we get significant fluid inertia during this period which has also been reported before \citep{Castrejon-Pita2015}, suggesting the domination of inertial forces over viscosity during the small time period. \textcolor{black}{The value of $u$ across the centerline has been shown in figure S4 (in supplementary) that also indicates high acceleration.} This is why equation~\ref{Equation:3.4} predicts the circulation values to the correct order. However, although viscosity is not dominant, it is important since the phenomenon takes place in the time scale similar to the order of $\tau_\mathrm{vis}$. Hence, to come to an exact solution we need to incorporate viscous effects to the equation that remains an open problem.}}
 
 We further observe a linear correlation (figure~\ref{fig:4}(g)) between $U_\mathrm {int}^*$ and the peak velocity ($U_\mathrm {peak}^*$ inside the droplet, as depicted in figure~\ref{fig:3}(b)) especially when these normalized velocities are $\leq1.1$, as highlighted through orange zone in  figure~\ref{fig:4}(g) \textcolor{black}{which represents later stages}. This proves $U_\mathrm {int}^*$ to be a suitable choice for the velocity scale, imposed earlier to explain vortex dynamics. Similar to $U_\mathrm{int}^*$, the peak velocity i.e.,  $U_\mathrm{peak}^*$ is also render non-dimensional using the ratio $(D_\mathrm d/t_\mathrm {pinch})$, where the $t_\mathrm{pinch}$ is the pinch-off time for the respective $\varphi$. This ratio has an increasing trend (figure S3 in supplementary data) with increasing $\varphi$. The data show excellent collapse among various values of $\varphi$ for lower values of velocity \textcolor{black}{(or in the later stages post pinch-off)}. The correlation becomes weaker for higher values with $U_\mathrm {int}^*$ being larger than $U_\mathrm {peak}^*$. \textcolor{black}{The deviation of the correlation from the $45^\circ$ line is because post pinch-off it is expected to have residual of the peak velocity within the bulk and the newly created surface motion will have a different velocity in the  early stages since the interface relaxation is dominated by capillary effects. However, in the later stages the influence of the interfacial motion is diffused within the droplet bulk, therefore it is expected that the two velocities are correlated.}
 
The interface motion is a resultant of complex force balance arising from capillary, shear (viscous) and pressure driven effects, and the displaced fluid (due to interface motion) within the droplet will diverge in the y-direction as well, the intensity of which depends on the amount of confinement or the droplet size, among other effects. Since these forces and the droplet sizes are dependent on $\varphi$, hence, the slope of the curves in figure~\ref{fig:4}(g) are seen to slightly vary with $\varphi$. But this variation is not significant in the present conditions. It is evident that the high surface tension force post pinch-off at the trailing end governs the vorticity and the peak velocity decay.

\subsection{Vortex Dynamics: Retracting and Advancing Ligament Vortex}

\begin{figure}
  \centerline{\includegraphics[width=0.7\columnwidth]{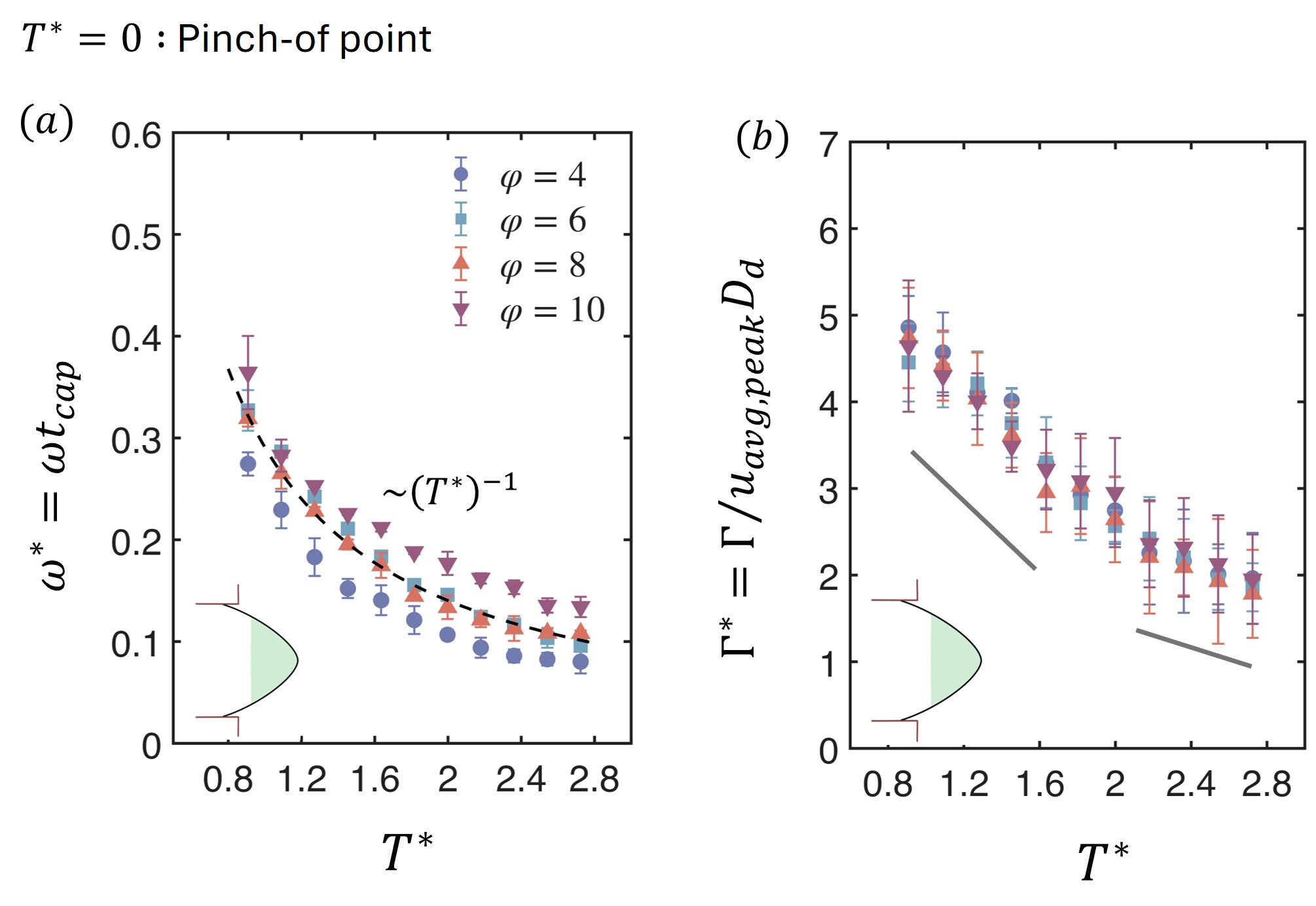}}
  \caption{(a) Evolution of $\omega^*$ for the retracting ligament vortex. The marked zone inside the droplet depicts the state of the droplet for which the data is taken. (b) Evolution of $\Gamma^*$ for the retracting ligament vortex. $u_\mathrm {avg,peak}$ represents the maximum of the average value of the velocity in the retracting ligament fluid.}
  
\label{fig:5}
\end{figure}

The bi-directional acceleration of the fluid during pinch-off results in a vortex in the retracting fluid as well. This vortex decays with the same scaling as observed for the post pinch-off vortex (figure~\ref{fig:5}(a)). This decay spans  a period of time ($\sim T^*=2.7$) beyond which the values begin to stagnate. The corresponding circulation ($\Gamma^*$ = $\Gamma$/$u_\mathrm {avg,peak} D_\mathrm d$, where $u_\mathrm {avg,peak}$ is the peak of the average velocity values) decays linearly with time, exhibiting a reduction in slope beyond a point, as shown in figure~\ref{fig:5}(b). An average value of velocity is used to render the circulation dimensionless for this case. \textcolor{black}{We observe a collapse of all the values when normalizing using the peak of average value and the steady state droplet diameter that varies with $\varphi$.}

\begin{figure}
  \centerline{\includegraphics[width=1\columnwidth]{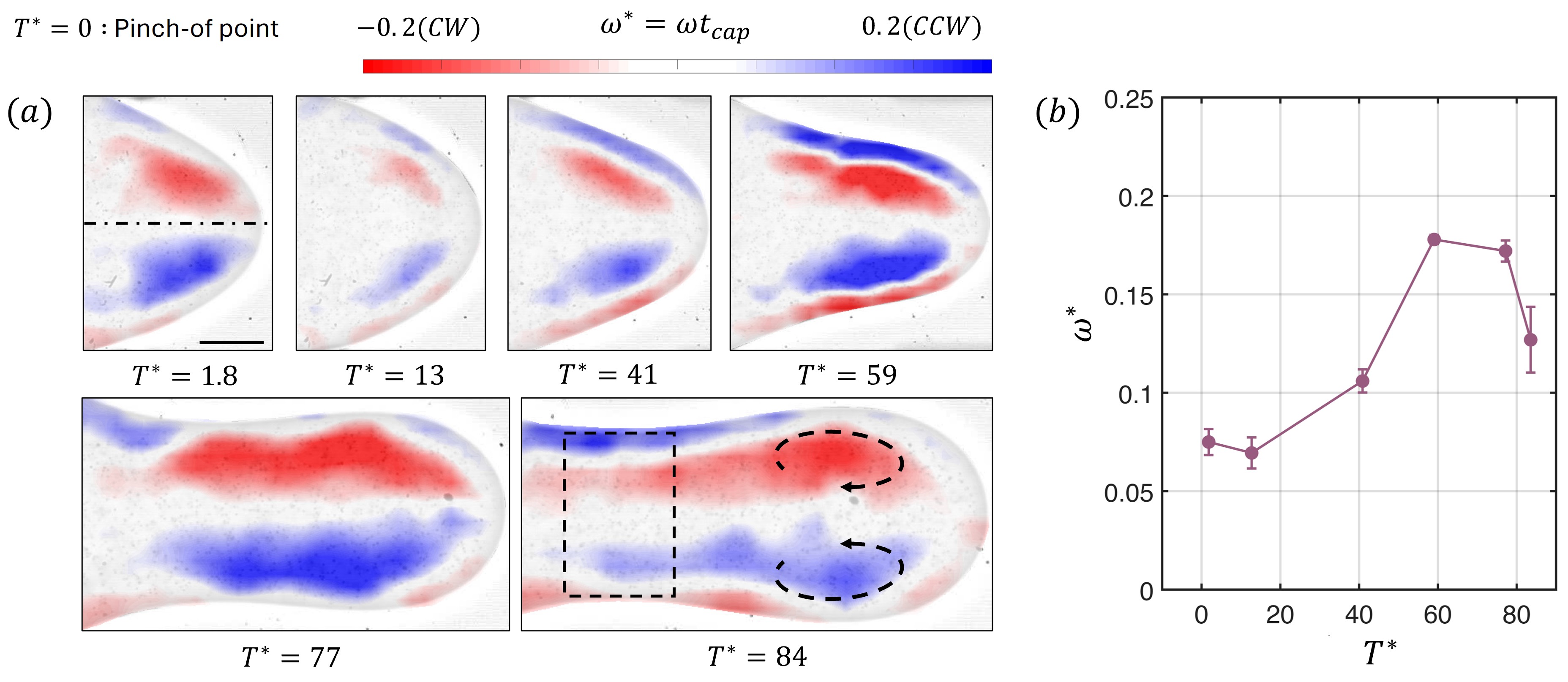}}
  \caption{(a) $\omega^*$ contours at different times for the advancing ligament vortex for $\varphi = 8$. $T^*=0$ marks the pinch-off point (b) The six $\omega^*$ values correspond to the vorticity contour plots in (a). The scale bar represents 50 $\mu m$.} 
\label{fig:6}
\end{figure}

Figure~\ref{fig:6}(a) depicts the vorticity contours for the transient advancing ligament, with $T^* = 0$ being the point of pinch-off for $\varphi = 8$. Figure~\ref{fig:6}(b) shows the $\omega^*$ for the corresponding contour plots shown in figure~\ref{fig:6}(a). These vorticity values are determined only for the central recirculating zone and not for the gradients near the interface. We discuss the phenomenon only for the region above the symmetry line, as demarcated in the first contour plot. As can be seen, beyond pinch-off, $\omega^*$ decreases slightly until $\sim$ $T^* = 13$ (this is continued from the retracted ligament vortex decay), beyond which there is a steep rise. Interestingly, during the rise, we observe an opposite vorticity gradient developing near the interface region (prominent at $T^* = 41, 59$). The interface changes its curvature due to pressure from the channel neck and continuous fluid (see the difference in curvature in figure~\ref{fig:6}(a): $T^*=1.8$ and $T^*=59$). The interface motion in a downward direction induces counter clockwise (\textit{CCW}) vorticity near the interface, consistent with what is discussed in the existing literature \citep{Song1999-kh,Terrington2020-pg}. This \textit{CCW} vorticity increases in strength from the first instance until the curvature changes sharply. However, this is impeded by the clockwise (\textit{CW}) rotating vortex generated during the pinch-off period. Notably, this retracting ligament vortex at the central region is still being fed by the fluid traveling along the interface (driven by shear) within dispersed fluid and curling back from the leading end, resulting in its vorticity enhancement. This outer \textit{CCW} vorticity near the interface is seen to decrease in strength once the dispersed phase completely fills the neck of the channel, i.e., when drag forces dominate over the capillary forces. We see a prominent circulating zone again ($T^* = 84$) in the bulging part of the dispersed phase fluid, i.e., the advancing ligament vortex. This is the same retracted vortex that exhibited a decay and then an increase in its vorticity. During this time the necking phase begins, and the post pinch-off droplet vortex starts to build near the neck region (highlighted with a dotted box at $T^* = 84$). Subsequently, the advancing ligament vortex in the bulk of the droplet decays and the post-pinch-off vortex comes into the picture.

\subsection {Stresses due to Post Pinch-Off Vortex}

 In this section, we examine the contours of the in-plane shear rate ($\epsilon$) and plot two different quantities: Area integral of stresses \(\int\epsilon,\eta_\mathrm {x},\eta_\mathrm {y}  dA\), and the maximum value of stresses, i.e., $\epsilon_\mathrm {max}$, $\eta_\mathrm {x,max}$  $\eta_\mathrm {y,max}$. 

\begin{figure}
  \centerline{\includegraphics[width=1\columnwidth]{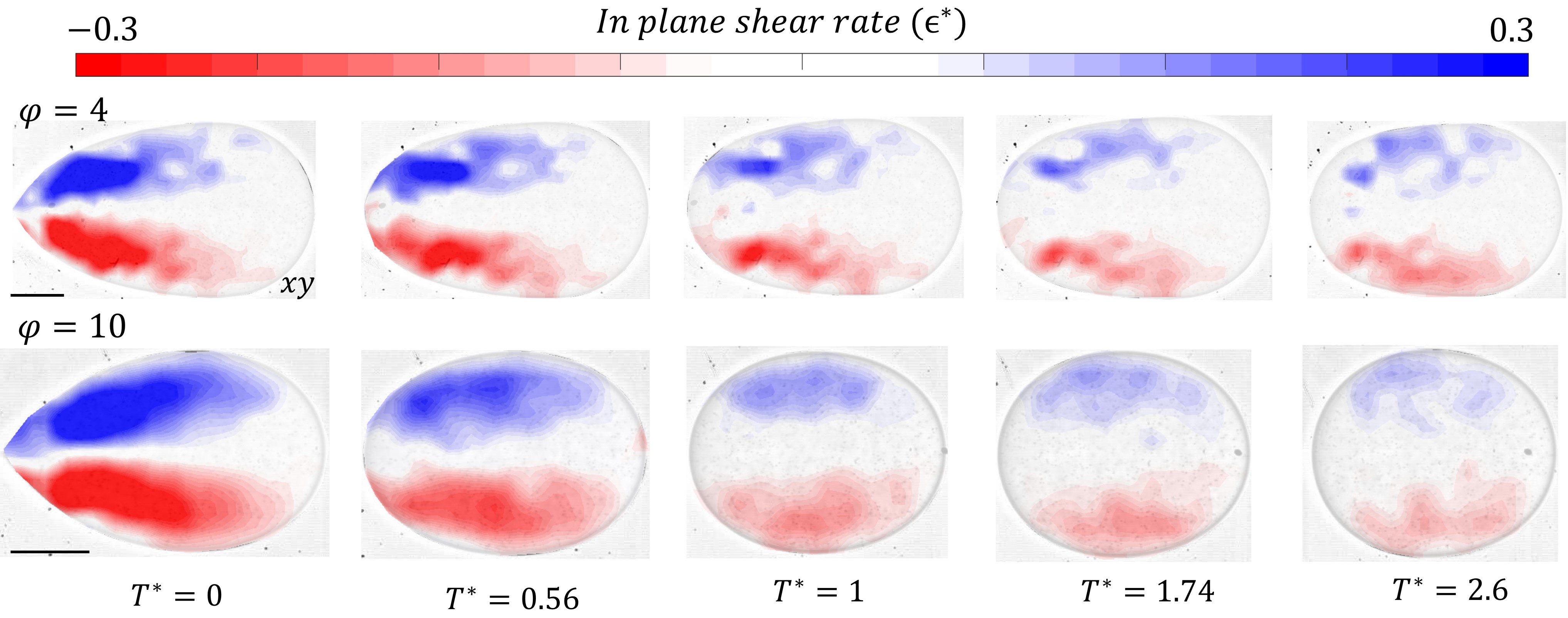}}
  \caption{Contour plots of $\epsilon^*$ at different times. The scale bar represents 50~$\mu m$}
\label{fig:7}
\end{figure}

Figure~\ref{fig:7} shows the contour plot for $\epsilon^*(=\epsilon t_\mathrm {cap})$ with $T^*$ for $\varphi= 4$ and 10. Similar to the $\omega^*$ contours in figure~\ref{fig:3}(a), we see that there is a decrease of $\epsilon^*$ with time and it covers a larger area for smaller droplets compared to larger droplets. \textcolor{black}{The area under the influence of $\epsilon^*$ demarcates the hot spots where a particle or a living organism will be subjected to intense shearing. As discussed in \citet{Tan2022}, the role of droplet size is crucial in bio-chemical processes taking place inside micro droplets. In droplets with smaller size ($\varphi$ = 10), a very significant portion is under the influence of shearing, which indicates a less favorable condition for entrapped living organism.} 
It is to be noted that the time of exposure to higher levels of $\epsilon^*$ will be less in this regime (post pinch-off vortex decay period); \textcolor{black}{however, the unbalanced surface tension post-pinch off rapidly accelerate both sides of the fluids that would act like an impact force \citep{Peregrine_Shoker_Symon_1990}} on any organism trapped in high shear zones and it has been shown that impact loadings can alter bacterial behavior significantly \citep{Hariharan2023}. Furthermore, figure~\ref{fig:8}(a) shows the variation of $\epsilon^*_{max}$  for each of the $\varphi$ values. It is seen that the values are slightly larger for smaller $\varphi$ compared to higher $\varphi$. The reason for this can be the size of the droplet, which is larger for lower $\varphi$, leading to a higher $D_\mathrm {d}/h$ ratio, indicating maximum activity in the observation plane. The other quantities shown in figure~\ref{fig:8}(b and c) for a single value of $\varphi = 6$ follow a similar decay and for all quantities it can be seen that $\epsilon^*$ is much higher compared to $\eta^*_\mathrm x(=\eta_\mathrm x t_\mathrm {cap})$ and $\eta^*_\mathrm y(=\eta_\mathrm y t_\mathrm {cap})$. Between $\eta^*_\mathrm x$ and $\eta^*_\mathrm y$, the linear shear in \textit{x} direction dominates. This is because the flow inside the droplet is driven by a shearing mechanism in the direction of the flow in \textit{x} direction. Also, the pinch-off stretches the droplet in the \textit{x} direction, generating its gradient in \textit{x} direction. The initial gradients are large enough to generate high impulsive loading inside the droplets.

\subsection {Stresses due to Retracted and Advancing Ligament Vortex}

\begin{figure}
  \centerline{\includegraphics[width=1\columnwidth]{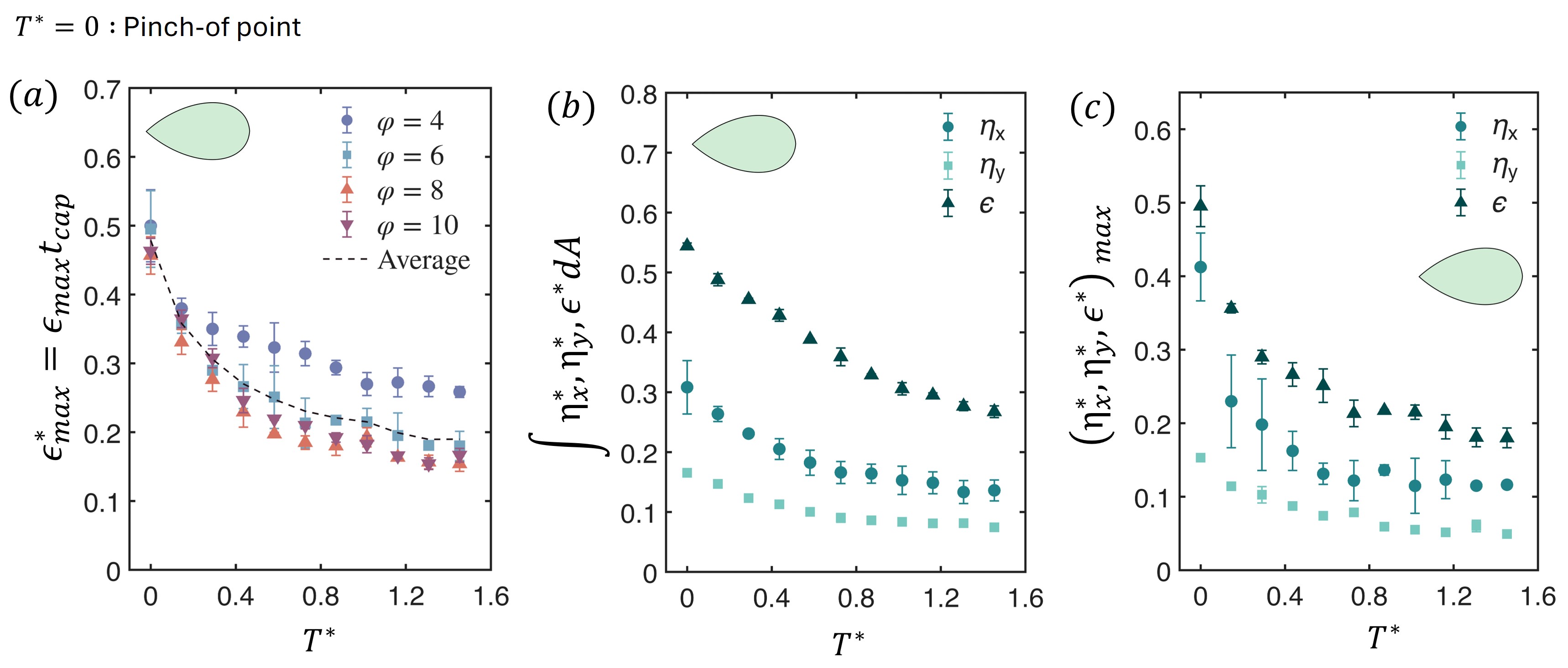}}
  \caption{(a) Variation of of $\epsilon^*_\mathrm {max}$ for $\varphi = 4-10$. \textcolor{black}{(b) Area integral normalized using $U_\mathrm{peak} D_d$ and (c) Maximum values of shear and extension rates for $\varphi = 6$.}}
\label{fig:8}
\end{figure}

\begin{figure}
  \centerline{\includegraphics[width=1\columnwidth]{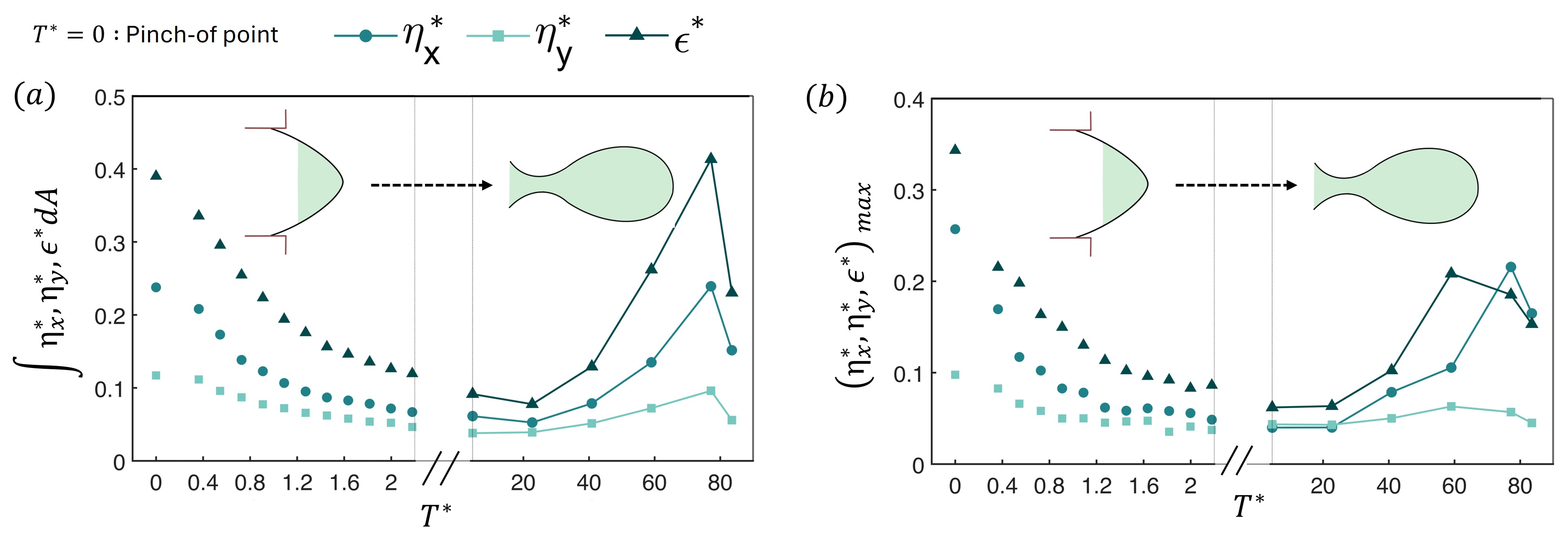}}
  \caption{ \textcolor{black}{Variation of $\epsilon^*$, $\eta^*_\mathrm x$ and $\eta^*_\mathrm y$ for retracted and advancing ligament phases (a) Area integral normalized by $U_\mathrm{peak} D_d$ (b) Maximum value.} The first part of each plot is for the retracted ligament. The other part of each plot represents the advancing ligament and corresponds to the same time as shown for vorticity plots in figure~\ref{fig:6}(a).}
\label{fig:9}
\end{figure}

 Corresponding to figure~\ref{fig:6}(a), where we depict the vorticity contours, we plot the different stress quantities in figure~\ref{fig:9}(a and b) for $\varphi=8$. Again,  $T^*=0$ represents the pinch-off point and the first part of the plots before the axis break represent the stresses during the decay of the retracting ligament vortex (the vorticity contours for the first part of the x axis is not presented whereas the other part can be correlated with the $T^*$ values in figure~\ref{fig:6}(a)). The initial decay in the retracted vortex is very similar to that in the pinch-off droplet, as was seen in figure~\ref{fig:8}(b, c). However, the extended part of the plot, where we show these quantities for longer $T^*$, indicates that the area integral of stress (figure~\ref{fig:9}(a)) increases nearly to the same level as during the pinch-off. This increase is due to the fact that different zones with velocity gradients are created with significantly high $\omega^*$ values in the advancing dispersed fluid. The most interesting aspect is the sustained presence of the retracted vortex with a decrease and subsequent increase in its $\omega^*$ value (approaching towards advancing ligament vortex), which results in large stresses. \textcolor{black}{The final dip seen in figure~\ref{fig:9} (a, b) is due to the decrease in the velocity gradients and stresses that occur near to the pinch-off ($T^*\sim 94$ ), when the vortex formed in the bulk of the droplet (as marked in figure~\ref{fig:6}, $T^* = 84$) starts to decay due to reduction in its feeding as a result of ligament thinning. We found no previous reference to the stresses developed in the advancing ligament, which nevertheless are very important, since this stress remains for a longer period compared to the post pinch-off stress in the droplet. }
 
\textcolor{black}{From a practical point of view, the stresses must be reduced inside the dispersed phase to provide habitable environment to the encapsulated cells \citep{Ma}. It should be understood that the particles or organisms remain under stresses right from the beginning of the cycle i.e., inside the remnant ligament, its advancement along the flow direction, during the pinch-off and finally in the steady state condition (as described in figure~\ref{fig:1}(b)).} Although not discussed here, the steady state of the droplet exhibits two counter rotating vortices of opposite sense \citep{Ma} compared to the post pinch-off vortex due to the action of the outer continuous fluid (figure~\ref{fig:1}(b)). The longer the organisms are entrapped, the longer they are subjected to stress. However, once the droplet attains a steady state, it is usually seen that the vorticity existing inside the droplet has relatively low magnitude depending on the $Q_\mathrm c$ value. So, a trade-off between the stress duration and stress magnitude occurs at different regimes of droplet formation process.

\section{Conclusions}

We present the vortex dynamics involved in pre, post and during the pinch-off process of a micro droplet inside a cross flow generator. \textcolor{black}{The pinch-off phenomenon is initially observed to depend on $\varphi$, whereas, the final stage is capillary-dominated. Consequently, the vorticity generated during pinch-off exhibited little variations with $\varphi$.} It is seen that the post pinch-off vortex decays as $\sim t^{-1}$ in the pinched off droplet and in the same manner in the retracting ligament. This is explained using the decay trend of the trailing end interface velocity $U^*_\mathrm {int}$, which comes from the high capillary forces acting at the trailing end of the droplet post pinch-off. \textcolor{black}{The collapse of $U^*_\mathrm {int}$ for different values of $\varphi$ is obtained through the velocity scale defined as $D_d/t_\mathrm{pinch}$. Using the same velocity scale, a collapse between $U^*_\mathrm {int}$ and $U^*_\mathrm {peak}$ is observed for smaller values of velocity with a linear correlation.}

Although, it is seen that the post pinch-off vortex decays rather quickly, it can be hypothesized to act like an impulsive loading on trapped organisms from an \textcolor{black}{application point of view}. The generation of velocity gradients during the entire pinch-off process in the dispersed phase is discussed in context of vorticity and stress fields. We predict the scale of maximum circulation $(\Gamma^*_\mathrm {peak})$ during the necking phase inside the forming droplet close to the obtained values using a slug flow approximation. Interestingly, the retracted vortex is sustained for a much longer time, with a decrease and subsequent increase in  $\omega^*$,  transforming into what we call the advancing ligament vortex before being pushed inside the bulk of the liquid droplet. This vortex is normally not discussed in literature, which potentially can create large stresses over longer time periods on living cells. The stress values exhibit a decay in the retracting ligament and then an increase inside the advancing ligament due to the strengthening of the vortex. The average stress values in the advancing ligament are seen to increase continuously until near the pinch-off time. 

The study explores how inherent \textcolor{black}{flow gradients} formed during the pinch-off process can generate significant stresses across varying magnitudes and time scales due to the vortical structures. \textcolor{black}{This when seen from a practical standpoint in biological processes like cell encapsulation, cytometry etc.,  potentially causes harmful effects on trapped organisms. In many practical scenarios, researchers aim to achieve smaller droplet sizes to minimize fluid volume and reduce channel size. However, this comes at the expense of an increased droplet surface area under the influence of shear stress, which must be carefully considered}. This area of research has been largely neglected by the fluid dynamics community, which significantly influences some very fundamental aspects of biological investigations. The steady state of the droplet generation process will be studied in future work.

\backsection[Acknowledgements]{S.J and S.J.R would like to thank the Prime Minister Research Fellowship (PMRF) for the financial support. C.T. would like to acknowledge the financial support from the Anusandhan National Research Foundation (ANRF) through the VAJRA Faculty scheme. S.B. would like to acknowledge the support from the Indian National Academy of Engineering (INAE) Chair professorship.}

\backsection[Declaration of interests]{The authors report no conflict of interest.}

\backsection[Author ORCIDs]{S. Jain, https://orcid.org/0000-0002-2756-8988; S. J. Rao, https://orcid.org/0000-0001-6539-5814; S. Mandal, https://orcid.org/0000-0001-8154-0989; C. Tropea, https://orcid.org/0000-0002-1506-9655; S. Basu, https://orcid.org/0000-0002-9652-9966}

\bibliographystyle{jfm}
\bibliography{jfm}

\section{Supplementary Material}

\begin{figure}
  \centerline{\includegraphics[width=0.4\columnwidth]{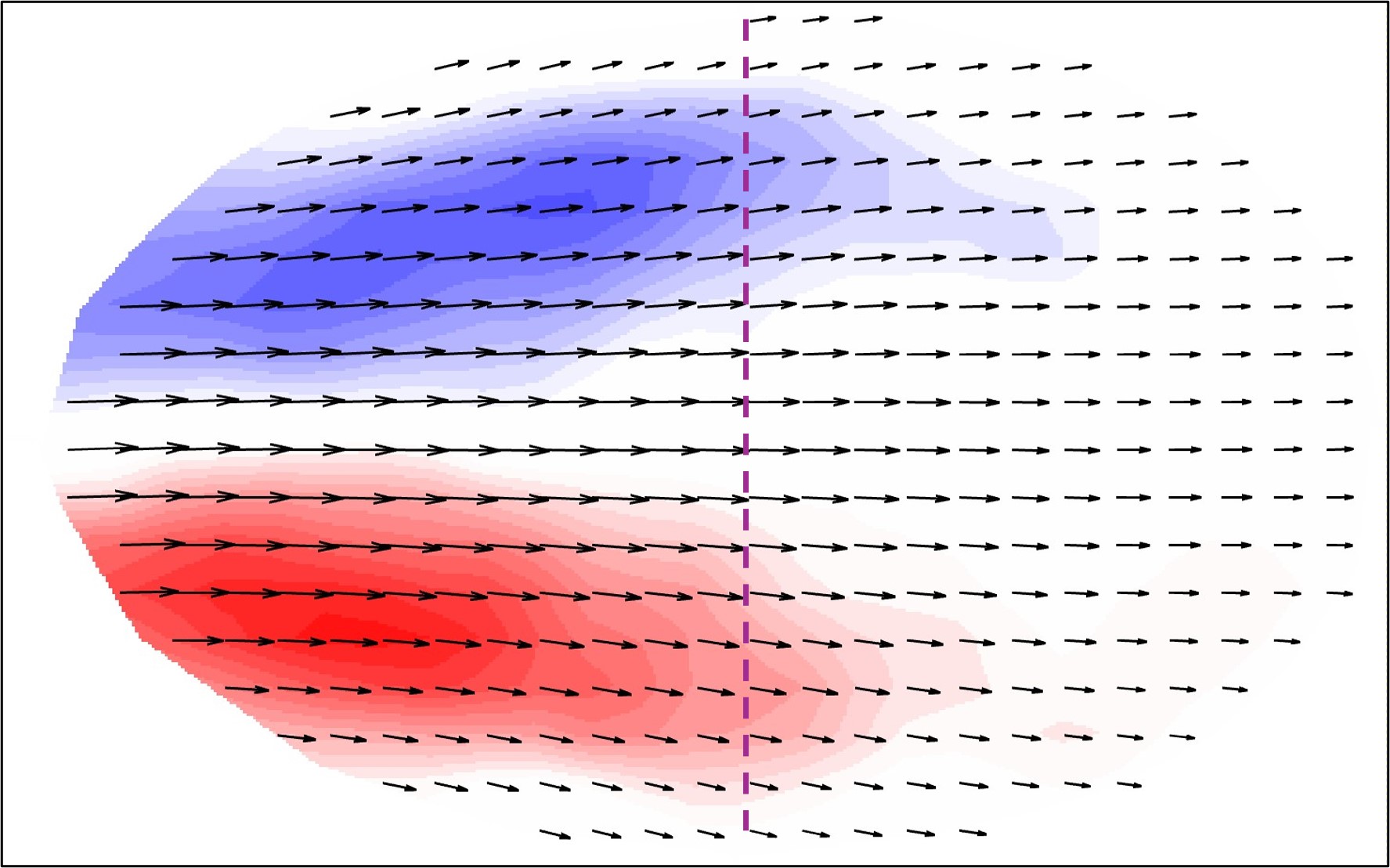}}
  \caption{\textbf{Figure S1:}The vector field in the lab frame of reference inside a pinched-off droplet overlaid over vorticity contours. The width of the droplet (marked by dashed line) ~ 125 $\mu m$, i.e., the gap between two vectors ~ 7 $\mu m$.}
\label{fig:S1}
\end{figure}

\begin{figure}
  \centerline{\includegraphics[width=0.45\columnwidth]{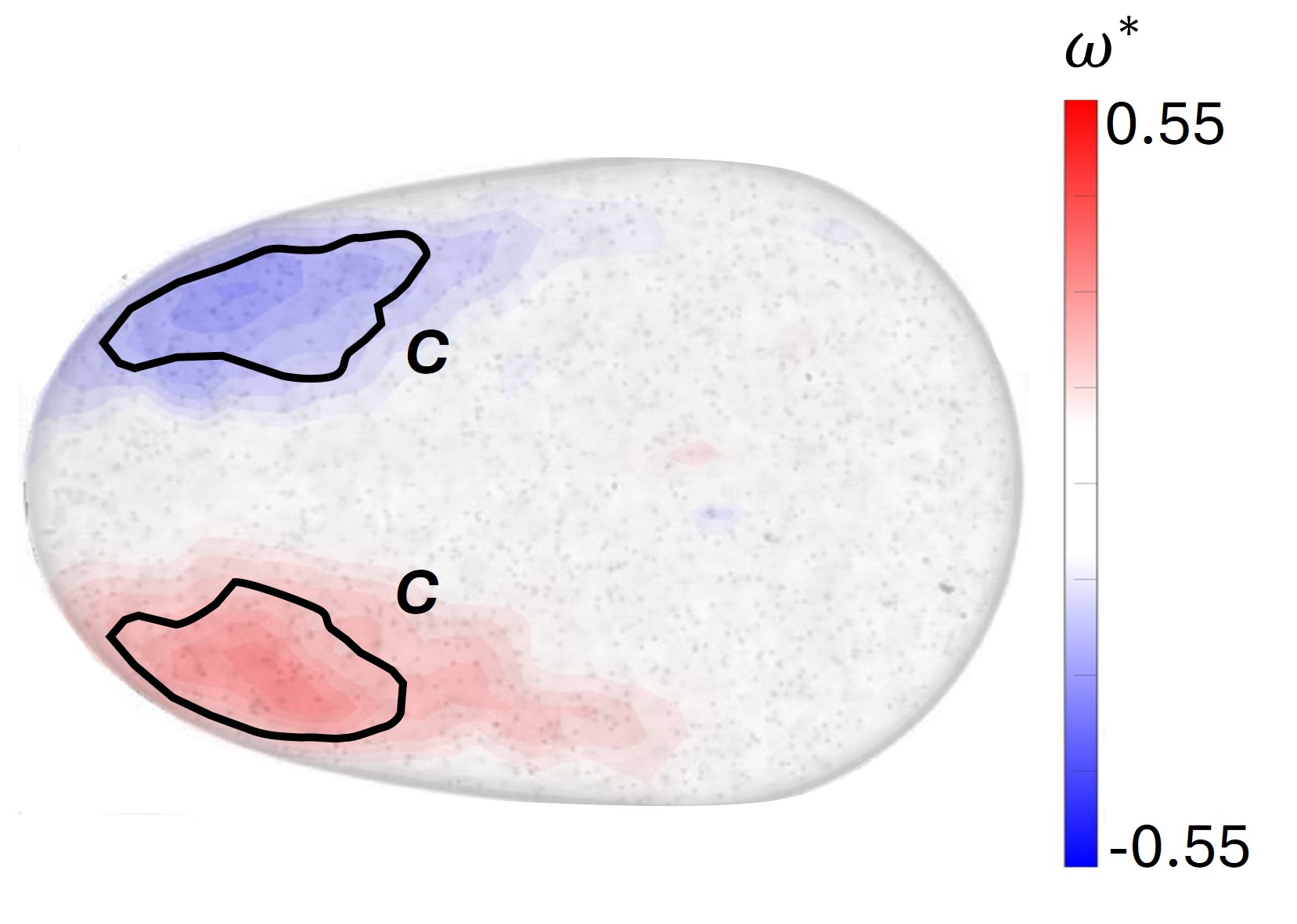}}
  \caption{\textbf{Figure S2}: The vorticity contour inside a pinched-off droplet is depicted with the vortex core region marked by C as detected using the $\lambda_{ci}$ method. }
\label{fig:S2}
\end{figure}

\begin{figure}
  \centerline{\includegraphics[width=0.85\columnwidth]{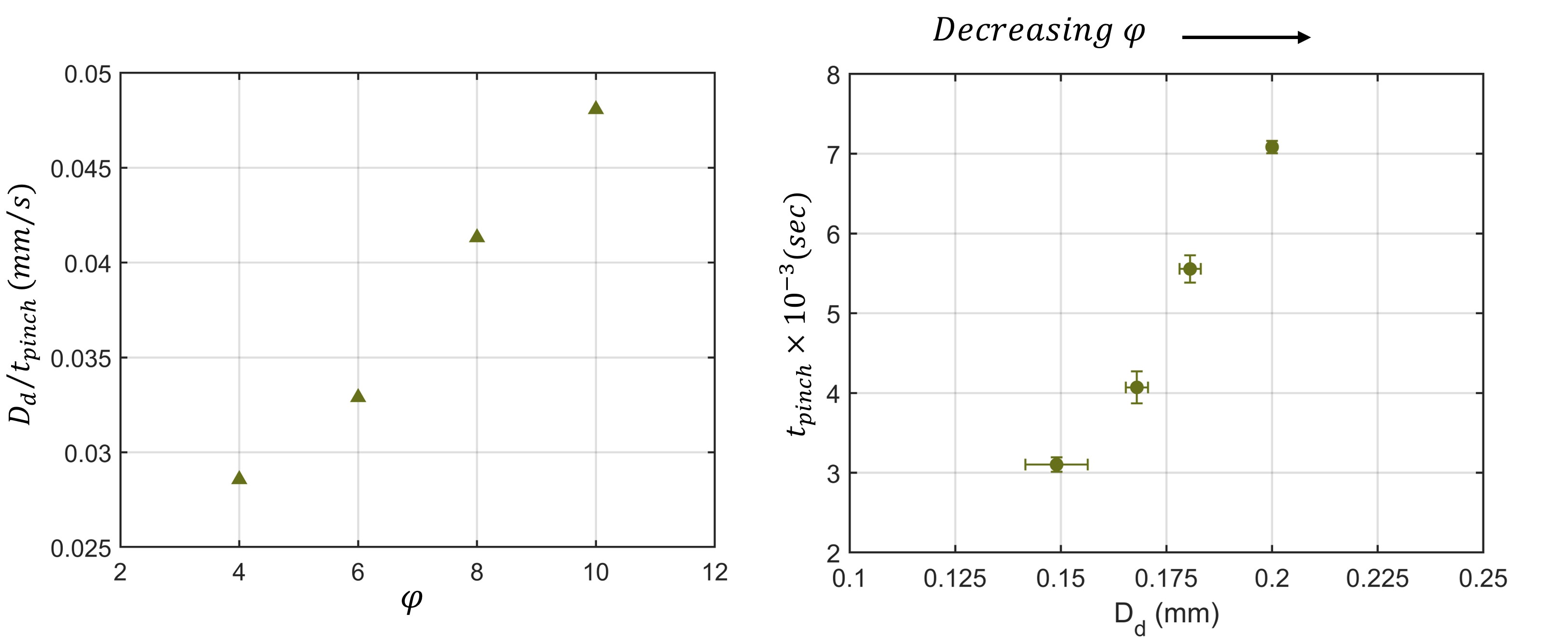}}
  \caption{\textbf{Figure S3}: (a) The variation of the velocity scale used to normalize the interfacial ($U_\mathrm{int}$) and peak velocity ($U_\mathrm{peak}$). (b) The variation of the ligament pinch-off time ($t_\mathrm{pinch}$) with the droplet diameter ($D_d$) in their dimensional form.}
\label{fig:S3}
\end{figure}

\begin{figure}
  \centerline{\includegraphics[width=0.7\columnwidth]{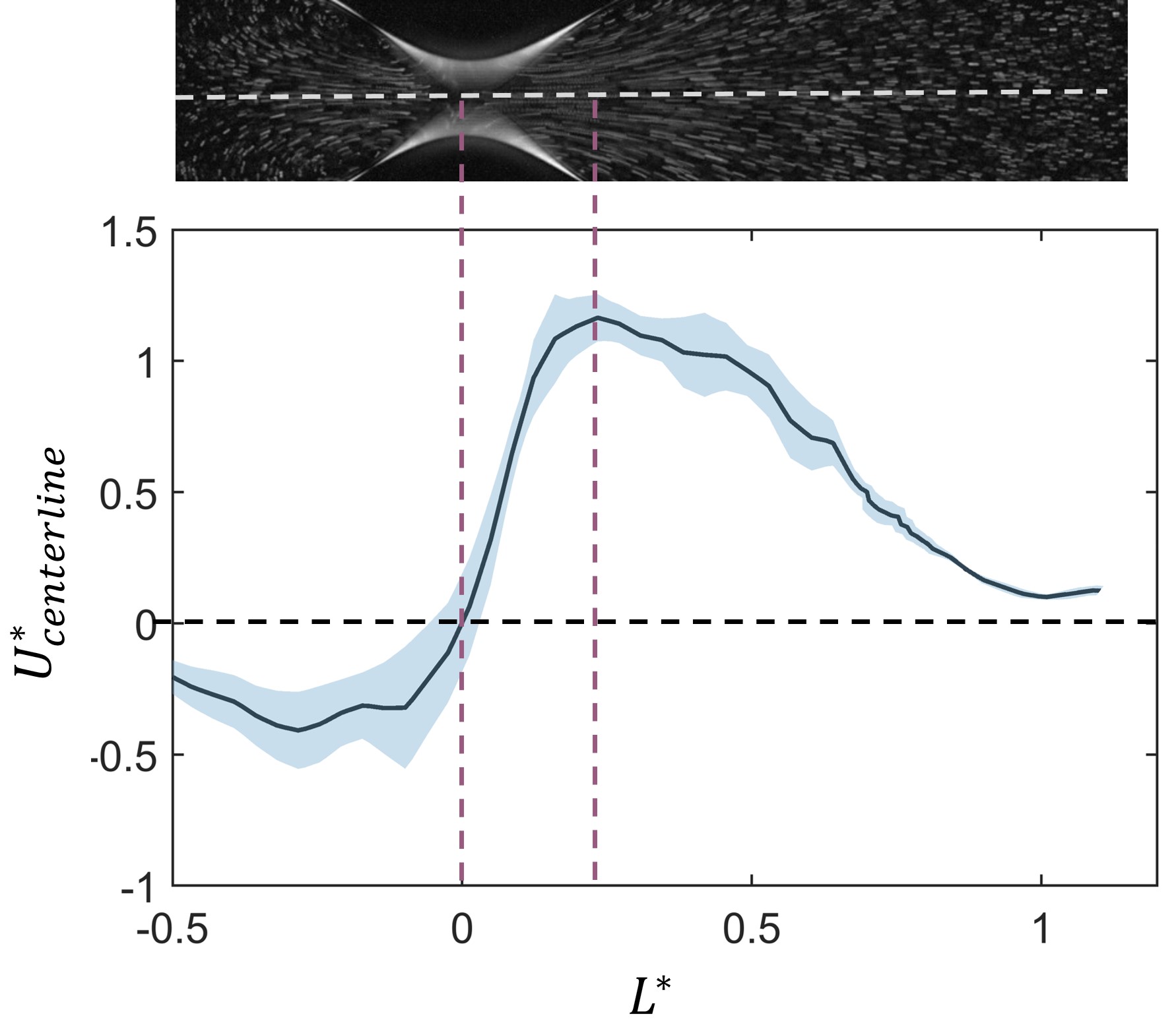}}
  \caption{\textbf{Figure S4:} The $x$ velocity ($u$) profile along the centerline.  This is obtained by assuming that the interface motion is negligible in $x$ direction and the dominant motion is in $y$ direction due to thinning of the ligament. The velocity is normalized using the time scale $D_d/t_{\mathrm{pinch}}$, and the distance along the line is normalized using $W_n$. The white dashed line depicts the centerline about which the velocity profile is plotted. The black dashed line represents $u = 0$ line, and the purple dashed lines represent the point of flow reversal and peak.}
\label{fig:S4}
\end{figure}

\end{document}